\documentclass[prl,aps,nofootinbib,twocolumn,preprintnumbers,10pt]{revtex4}
\usepackage{amsmath,amssymb,graphicx,epsfig,hyperref,breakurl,color,bm,subfigure,slashed,lipsum,multirow}

\allowdisplaybreaks[4]

\begin{document}

\title{Establishing $CP$ Violation in $b$-Baryon Decays}
%\title{ 
%$CP$ violation of $b$-baryon decays in perturbative QCD }

\author{ Jia-Jie Han$^1$, 
Ji-Xin Yu$^1$~\footnote{Corresponding author: yujx18@lzu.edu.cn},
Ya Li$^2$~\footnote{Corresponding author: liyakelly@163.com},
Hsiang-nan Li$^3$~\footnote{Corresponding author: hnli@phys.sinica.edu.tw},
Jian-Peng Wang$^1$~\footnote{Corresponding author: wangjp20@lzu.edu.cn}, 
Zhen-Jun Xiao$^4$~\footnote{Corresponding author: xiaozhenjun@njnu.edu.cn},
Fu-Sheng Yu$^1$~\footnote{Corresponding author: yufsh@lzu.edu.cn} }

\affiliation{
$^1$Frontiers Science Center for Rare Isotopes, and School of Nuclear Science and Technology, Lanzhou University, Lanzhou 730000,   People’s Republic of China \\
$^2$Department of Physics, College of Sciences, Nanjing Agricultural University, Nanjing 210095, People’s Republic of China
\\
$^3$Institute of Physics, Academia Sinica, Taipei, Taiwan 115, Republic of China
\\
$^4$Department of Physics and Institute of Theoretical Physics, Nanjing Normal University, Nanjing 210023, People’s Republic of China}

\begin{abstract}
It is a long-standing puzzle why the {\it CP} violation (CPV) in the baryon system has not yet been definitively established as in the meson one.
We demonstrate that individual partial-wave CPV in the $\Lambda_b\to p\pi^-$ and $pK^-$ decays can exceed $10\%$, but the destruction between the partial waves (the suppression by the small partial-wave weight) results in a small net direct CPV in the former (the latter) as measured currently. Our finding highlights the different dynamics responsible for CPVs in baryon and meson decays. 
We propose to probe the CPV observables associated with the angular distributions of the $\Lambda_b\to pa_1(1260)$, $pK_1(1270)$ decay products, which are large enough for being identified experimentally. 
\end{abstract}

%The above observation is supported by the first full QCD calculation of two-body hadronic $\Lambda_b$ baryon decays with controllable uncertainties in the perturbative QCD formalism. 

\pacs{xxxx}

\maketitle

\textit{Introduction}\textemdash{}
The {\it CP} violation (CPV) plays a crucial role in explaining the matter-antimatter asymmetry in the Universe.
CPVs in the standard model (SM) originate from an irreducible phase in the Cabibbo-Kobayashi-Maskawa (CKM) matrix, which is, however, much smaller than those required by the matter-antimatter asymmetry. 
It is important to explore new CPV sources from physics beyond the SM.
Currently, the experimentally observed CPVs are all in the meson systems, such as $K$ \cite{Christenson:1964fg}, $B$ \cite{BaBar:2001ags,Belle:2001zzw}, and $D$ \cite{LHCb:2019hro} mesons.
However, CPVs in baryon systems have not yet been firmly established \cite{BESIII:2021ypr,BESIII:2018cnd,LHCb:2017hwf,LHCb:2016yco,LHCb:2018fly,LHCb:2018fpt,LHCb:2019oke,LHCb:2019jyj,LHCb:2024yzj,LHCb:2024iis}.
Given that the visible matter in the Universe is predominantly composed of baryons, it is natural to investigate CPVs in baryon decays.
Recently, the LHCb Collaboration reported evidence of CPV in $\Lambda_b^0\to \Lambda^0K^+K^-$ \cite{LHCb:2024yzj}, which motivates timely theoretical studies of baryonic CPV.

Unlike mesons, baryons cannot undergo mixing and, thus, exhibit only CPVs in decays, known as direct CPVs.
Bottom hadron decays allow large direct CPV at the order of $10\%$, in principle, owing to the relatively large CKM weak phase between the first and third generations of quarks. 
This expectation has been confirmed in {\it B} meson decays with the measured direct CPVs $(-8.31\pm0.31)\%$ for $B^0\to K^+\pi^-$, $(22.4\pm1.2)\%$ for $B_s^0\to K^-\pi^+$, $(-31.4\pm3.0)\%$ for $B^0\to \pi^+\pi^-$, and $(16.2\pm3.5)\%$ for $B_s^0\to K^+K^-$ \cite{ParticleDataGroup:2024cfk}.
By contrast, CPVs in the baryon system have not been identified. For example, the measured direct CPVs \cite{LHCb:2024iis}
\begin{equation}\label{eq:AcpExp}
\begin{split}
	A_{CP}^{\rm dir}(\Lambda_b^0\to p\pi^-)&=(0.2\pm 0.8\pm 0.4)\%,\\
	A_{CP}^{\rm dir}(\Lambda_b^0\to pK^-)&=(-1.1\pm 0.7\pm 0.4)\%,
\end{split}
\end{equation}
are compatible with null asymmetries within the precision of $1\%$. That is, CPVs in $\Lambda_b$ baryon decays are much lower than those in similar {\it B} meson decays, although both occur through the $b\to u\bar{u}q$ transitions, $q=d,s$. Why baryon CPVs are so small and have not been established two decades after the observation of meson CPVs remains a puzzle in particle physics.

The dynamics in baryon and meson processes differ dramatically.
The half-integer spin of baryons induces at least two partial-wave amplitudes, such as $S$- and $P$-wave amplitudes in $\Lambda_b\to p\pi^-$, $pK^-$, while there is only one partial-wave amplitude in $B$ meson decays into two pseudoscalars. 
The analysis of the former is more intricate because of additional quarks involved in baryon modes, and at least two hard gluons to share momentum transfer, which may enhance higher-power contributions and modify the behavior of power expansions \cite{Wang:2011uv,Han:2022srw}. 
There also exist more topological diagrams, including color-commensurate $W$-emission and $W$-exchange diagrams, which provide abundant sources of strong phases required for direct CPVs.
A precise evaluation of strong phases in these topological diagrams poses a theoretical challenge. 
If the dynamical origin disparate from meson decays is not understood, CPVs in baryon processes cannot be predicted accurately.

%The sign of strong phases in different partial waves may be opposite to each other.
%due to the weak interactions with the vector and axial-vector currents. 

We will demonstrate in a full QCD calculation that CPVs in partial waves of bottom baryon decays can be as large as those in $B$ meson decays actually, greater than $10\%$, but the destruction between partial waves (the suppression by the small $P$-wave weight) turns in a small net CPV for the $\Lambda_b\to p\pi^-$ ( $\Lambda_b\to pK^-$) decay. 
We further analyze the $\Lambda_b\to p\rho^-, pK^{\ast -}, pa_1^-(1260)$ and $pK_1^-(1270)$ modes, whose partial-wave CPVs, overall speaking, also reach 10$\%$, but cancel each other, giving rise to small total direct CPVs. 
Our study manifests the different dynamics responsible for CPVs in bottom baryon and meson decays and suggests to detect the former through partial-wave-related CPV observables.
In particular, the magnitudes of CPVs via the up-down asymmetries in  the angular distributions of the $\Lambda_b\to pa_1(1260)$, $pK_1(1270)$ decay products are higher than $20\%$, granting a great chance to establish baryon CPVs.

%%%%%%%%%%%%%%%%%%%%%%%%%%%%%%%%%%
\textit{CP asymmetries in $\Lambda_b\to p\pi^-$, $pK^-$}\textemdash{}
A baryon decay amplitude, such as the one for $\Lambda_b\to p h$ with $h=\pi^-$ or $K^-$, is decomposed into
\begin{equation}\label{eq:amp=S+P}
	\mathcal{A}(\Lambda_b\to ph)=i\bar{u}_p (S+P\gamma_5)u_{\Lambda_b},
\end{equation}
where $u_p$ ($u_{\Lambda_b}$) is the proton ($\Lambda_b$) spinor, and $S$ ($P$) denotes the parity-violating $S$-wave (parity-conserving $P$-wave) amplitude.
The direct CPV is defined as
\begin{equation}
	\begin{split}
		&A_{CP}^{\rm dir}(\Lambda_b\to ph)\equiv \frac{\Gamma(\Lambda_b\to ph) - \bar \Gamma(\bar{\Lambda}_b\to \bar{p}\bar{h})}{\Gamma(\Lambda_b\to ph) + \bar \Gamma(\bar{\Lambda}_b\to \bar{p}\bar{h})},
		\label{eq:dirCPV}
	\end{split}
\end{equation}
with $\Gamma\propto |S|^2+\kappa|P|^2$, $\bar \Gamma\propto |\bar S|^2+\kappa|\bar P|^2$, and $\kappa=[(m_{\Lambda_b}-m_p)^2-m_h^2]/[(m_{\Lambda_b}+m_p)^2-m_h^2]\approx0.51$, $\bar S$ and $\bar P$ being the $CP$-conjugate amplitudes.
We also define the partial-wave CPVs:
\begin{equation}
	\begin{split}
		A_{CP}^{S\text{-wave}}\equiv \frac{|S|^2-|\bar{S}|^2}{|S|^2+|\bar{S}|^2}
        %=\frac{-2r_1 \sin\Delta\phi \sin\Delta\delta_1}{1+r_1^2+2r_1\cos\Delta\phi \cos\Delta\delta_1}
        ,~~~~
		A_{CP}^{P\text{-wave}}\equiv \frac{|P|^2-|\bar{P}|^2}{|P|^2+|\bar{P}|^2}
        %=\frac{-2r_2 \sin\Delta\phi \sin\Delta\delta_2}{1+r_2^2+2r_2\cos\Delta\phi \cos\Delta\delta_2}
        .
		\label{eq:partialCPV}
	\end{split}
\end{equation}
The total direct CPV can be expressed as the weighted sum of the partial-wave CPVs:
\begin{equation}
    A_{CP}^{\rm dir}= \kappa_S A_{CP}^{S\text{-wave}} + \kappa_P A_{CP}^{P\text{-wave}}, 
    \label{eq:ACP=S+P}
\end{equation}
with the coefficients $\kappa_S=|S|^2/(|S|^2+\kappa r_{CP}|P|^2)$ and $\kappa_P=\kappa|P|^2/(|S|^2/r_{CP}+\kappa|P|^2)$, where $r_{CP}=(1+A_{CP}^{S\text{-wave}})/(1+A_{CP}^{P\text{-wave}})$.

%\textcolor{red}{assuming small partial-wave CP violations}.

CPVs are generated by the interference between the contributions of the {\it tree} operators $\mathcal{T}$ and the {\it penguin} operators $\mathcal{P}$. The $S$- and $P$-wave amplitudes are then decomposed into
\begin{equation}
\begin{split}
	S=\lambda_{\mathcal{T}}|S_{\mathcal{T}}|e^{i\delta^S_{\mathcal{T}}} + \lambda_{\mathcal{P}}|S_{\mathcal{P}}|e^{i\delta^S_{\mathcal{P}}},\\
	P=\lambda_\mathcal{T}|P_\mathcal{T}|e^{i\delta^P_\mathcal{T}} + \lambda_\mathcal{P}|P_\mathcal{P}|e^{i\delta^P_\mathcal{P}},
\end{split}\label{str}
\end{equation}
respectively, where $\delta$'s represent the strong phases and $\lambda_\mathcal{T}=V_{ub}V_{ud(s)}^*$ and $\lambda_\mathcal{P}=V_{tb}V_{td(s)}^*$ are the products of the CKM matrix elements for the $\Lambda_b\to p\pi^-(K^-)$ decay.
The $b\to d\bar u u$ ($b\to s\bar u u$) transitions are dominated by tree (penguin) contributions with $|V_{td}^*V_{tb}/V_{ud}^*V_{ub}|(\alpha_s/\pi)\sim0.1$ [$|V_{ts}^*V_{tb}/V_{us}^*V_{ub}|(\alpha_s/\pi)\sim2$]. 
Considering the above ratios, we expect that the CPVs in these modes are of the order of $10\%$.  
The small measured CPVs in $\Lambda_b\to p\pi^-$, $pK^-$ hint either that the strong-phase differences between the tree and penguin components in Eq.~(\ref{str}) diminish in the dominant partial wave, or that the CPVs in the $S$- and $P$-wave amplitudes cancel in Eq.~(\ref{eq:ACP=S+P}). 
%We will elucidate that it is the latter which accounts for the tiny CPVs in Eq.~(\ref{eq:AcpExp}). 

%and the subscript $\mathcal{T}$ ($\mathcal{P}$) labels the {\it tree} ({\it penguin}) contribution. 
%the weak phases $\phi$ from the CKM matrix take the same values for the $S$- and $P$-waves, 

%
\begin{figure}[!tbhp]
	\centering
	\includegraphics[width=1.0\linewidth]{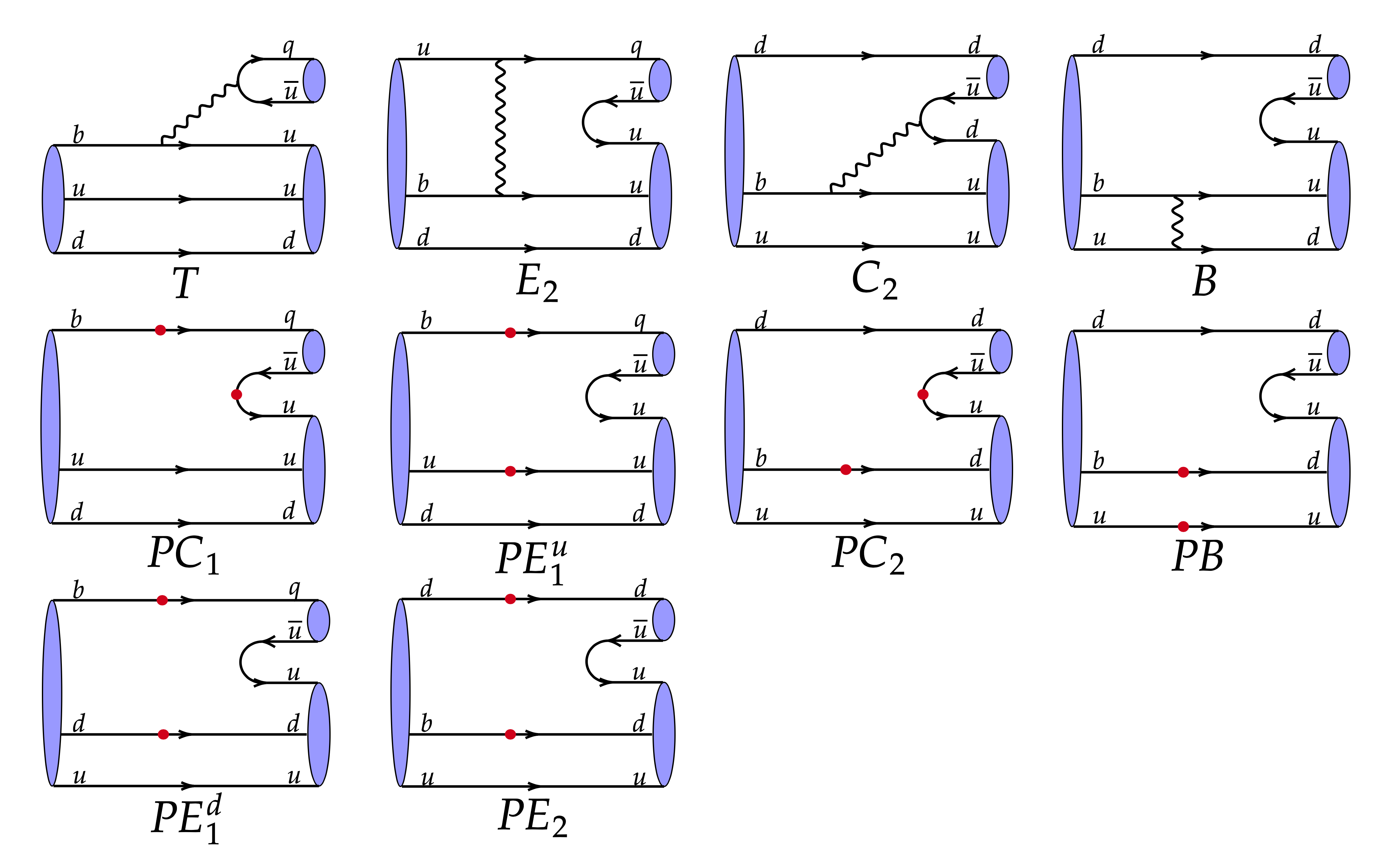}
	\caption{Topological diagrams contributing to the $\Lambda_b\to ph$ decays with the light quark $q=d, s$, where the wavy line represents a $W$ boson and  
    the red dots represent the insertion of the penguin operators.}
	\label{fig:topo}
\end{figure}

%The first row collects the tree diagrams with 
%The second and third rows collect the penguin diagrams with 

%and include all strong dynamics
%with the meson being formed by the quark decayed from light quark; 
%with the meson being formed by the light spectator quark.

Topological diagrams, useful for assessing hadronic heavy-flavor decays  \cite{Chau:1982da,Chau:1986jb,Chau:1995gk},
are classified according to weak currents and quark flows.
The topological diagrams for the $\Lambda_b\to ph$ decays, $h$ being a meson formed by the quarks $\bar u d$ or $\bar u s$, are displayed in Fig.~\ref{fig:topo}. 
There are four tree diagrams: the color-allowed external $W$-emission $T$, the $W$-exchange  $E_2$, the color-commensurate $W$-emission $C_2$, and the bow-tie $W$-exchange $B$. 
The other diagrams $PC_1$, $PC_2$, $PE_1^u$, $PE_1^d$, $PE_2$, and $PB$ are from the QCD and electroweak penguin operators.
The diagrams $T$, $E_2$, $PC_1$, $PE_1^u$, and $PE_1^d$ contribute to the modes with the final states ${\bar{u}d}$ and ${\bar{u}s}$, while $C_2$, $B$, $PC_2$, $PB$, and $PE_2$ contribute only to the former. 

%For convenience, we do not associate the CKM matrix elements with the topological diagrams. 

%The penguin contributions to the $S$-wave amplitude are dominated by $PC_1$, $S_{PC_1}\propto a_4+R_h a_6$, due to the chiral enhancement, while the $PC_1$ contribution to the $P$ wave, $P_{PC_1}\propto a_4-R_h a_6$, is suppressed by the destruction between the two terms, where $a_{4,6}$ are the Wilson coefficients and the mass ratio $R_h=2m_h^2/[m_b(m_u+m_q)]\approx 1$  \cite{Hsiao:2014mua,Zhu:2018jet} with the light meson ($b$ quark, $u$ quark, light quark) mass $m_h$ ($m_b$, $m_u$, $m_q$).

The diagram $T$, dominating the heavy-quark expansion \cite{Leibovich:2003tw}, yields the $S$-wave amplitude $S_{\mathcal{T}}=a_1(m_{\Lambda_b}-m_p)f_1(m_h^2)$ and the $P$-wave amplitude $P_{\mathcal{T}}=a_1(m_{\Lambda_b}+m_p)g_1(m_h^2)$ \cite{Cheng:1996cs}, where $a_1$ is the Wilson coefficient, and $f_1\approx g_1$ \cite{Han:2022srw,Manohar:2000dt,Khodjamirian:2011jp,Faustov:2016yza,Detmold:2015aaa} are the $\Lambda_b\to p$ transition form factors. 
Therefore, the tree contributions to the $S$- and $P$-wave amplitudes are of the same sign, $S_{\mathcal{T}}\approx P_{\mathcal{T}}$.
To acquire the relative sign between the $S$- and $P$-wave penguin contributions, we convolute a topological diagram naively with the effective weak operators and the dominant combination of distribution amplitudes (DAs) of initial and final states. 
For example, the $PC_2$ and $PE_1^u$ amplitudes are expressed as $\bar{u}_p (1+\gamma_5) (\gamma_5\slashed{p}_{\pi}) (\slashed{v}_{\Lambda_b}\gamma_5) \slashed{p}_p (1-\gamma_5)u_{\Lambda_b}$ and $(\bar{u}_p\gamma_\lambda)(\slashed{v}_{\Lambda_b}\gamma_5)(1+\gamma_5)(\slashed{p}_{\pi}\gamma_5)(\gamma_\lambda \slashed{p}_p)(1-\gamma_5)u_{\Lambda_b}$, respectively, where the weak vertex assumes the $(S-P)(S+P)$ form, and the pion momentum $p_\pi$, the proton momentum $p_p$ and the unit vector $v_{\Lambda_b}$ along the minus direction in light-cone coordinates are associated with the twist-2 pion, twist-4 proton and twist-4 $\Lambda_b$ DAs \cite{Han:2022srw}, respectively. It is trivial to simplify the expressions into $\bar{u}_p(1-\gamma_5)u_{\Lambda_b}$,
whose comparison with Eq.~(\ref{eq:amp=S+P}) then implies $S_{PC_2}\approx - P_{PC_2}$ and $S_{PE_1^u}\approx - P_{PE_1^u}$, i.e., the cancellation between the $S$- and $P$-wave amplitudes and small total direct $CP$ asymmetries. 
Such cancellation does not appear in $B$ meson decays into two pseudoscalar mesons, which involve only a single partial wave and no $PC_2$ and $PE_1^u$ diagrams.

The above crude argument by means of the structure of topological diagrams needs to be justified by comprehensive QCD calculations. 
Three popular theoretical approaches have been developed to study hadronic $B$ meson decays, known as the QCD factorization \cite{Beneke:1999br,Beneke:2000ry}, the soft-collinear-effective theory \cite{Bauer:2000yr,Bauer:2001yt,Bauer:2002nz}, and the perturbative QCD (PQCD) factorization\cite{Keum:2000wi,Lu:2000em,Keum:2000ph}. 
The PQCD approach is based on the $k_T$ factorization theorem and free of end-point singularities, in which
the factorizable and nonfactorizable emission, $W$-exchange, and annihilation diagrams are all calculable.
CPVs of two-body hadronic $B$ meson decays have been successfully predicted in PQCD \cite{Keum:2000wi,Lu:2000em,Keum:2000ph}.
Recently, this formalism was extended to the $\Lambda_b\to p$ transition form factors with reasonable high-twist hadron DAs, and the results agree with those from lattice QCD and other nonperturbative methods \cite{Han:2022srw}. Exclusive heavy baryon decays can, thus, be analyzed systematically in this framework.

%\begin{equation}
%	\begin{split}
%		\mathcal{A}(\Lambda_b & \to ph)=\int_0^1[dx] [dx^\prime] dy \int [d\bm{b}] [d\bm{b^\prime}] d\bm{b_{q}}\\
%		&H([x],[x^\prime],y,[\bm{b}],[\bm{b^\prime}],\bm{b_{q}},\mu) S_t([x],[x^\prime],y)\\ &\phi_{\Lambda_b}([x],[\bm{b}],\mu) \phi_p([x^\prime],[\bm{b^\prime}],\mu) \phi_{h}(y,\bm{b_{q}},\mu)\\
%		&e^{-S_{\Lambda_b}([x],[\bm{b}])}  e^{-S_{p}([x^\prime],[\bm{b^\prime}])}  e^{-S_{h}(y,[\bm{b_q}])}.
%	\end{split}
%\end{equation}

%The exchange of two hard gluons is necessary for $H$ at leading order in $\alpha_s$ to ensure the two light spectator quarks in the $\Lambda_b$ baryon to form the energetic final state. A typical diagram responsible for the $\Lambda_b\to p\pi^-$ decay is displayed in Fig.~\ref{fig:feyn-Cprime}. 

We compute the contributions from all diagrams to the $\Lambda_b\to p\pi^-$, $pK^-$ decays in the PQCD approach (see
Supplemental Material~\cite{supplement}), where an amplitude is written as a convolution of hard kernels and hadron DAs. It is worth mentioning that several topological diagrams, especially the nonfactorizable ones from $PC_2$ and $PE_1^u$, are evaluated for the first time. 
The hard kernel contains the weak vertices and two off-shell gluons. The DAs for the $\Lambda_b$ baryon, proton, and pesudoscalar mesons are input from Refs.~\cite{Ball:2008fw,Bell:2013tfa,Braun:2000kw,Braun:2006hz,Ball:2004ye,Ball:2006wn}. We summarize the predictions in Table~\ref{tab:topo}, where $\mathcal{T}$ and $\mathcal{P}$ are the total tree and penguin amplitudes, respectively, and only the central values are given for clarity.
It is seen that the imaginary $S$- and $P$-wave amplitudes from the $T$ diagrams are of the same sign, but those from the dominant penguin diagrams $PC_2$ and $PE_1^u$ are opposite in sign, consistent with our argument.  

\begin{table}[tbhp]
	\centering
	\renewcommand{\arraystretch}{1.1}
	\caption{Results of the topological amplitudes for the  $\Lambda_b\to p\pi^-$, $p K^-$ decays in unit of $10^{-9}$, which do not include the CKM matrix elements.}
	\label{tab:topo}
	\begin{tabular}{c@{\extracolsep{\fill}}cc|cc}
		\hline
		\hline
		~ Amplitudes~  &  ~ Re($S$)~  & ~ Im($S$)~  & ~Re($P$)~  & ~ Im($P$)~ \\
		\hline\hline
        \multicolumn{5}{c}{$\Lambda_b\to p\pi^-$}\\
        \hline
        $T$ &   701.19 &   $-$51.38 &   967.54 &  $-$265.17\\
        $C_2$ & $-$26.61 &    12.43 &   $-$41.51 &     0.14\\
        $E_2$ & $-$55.01 &   $-$38.14 &   $-$36.23 &    62.89\\
        $B$ &    $-$4.00 &     9.60 &   $-$12.73 &   $-$19.93\\%\hline
        Tree $\mathcal{T}$ &  615.57 &   $-$67.49 &   877.08 &  $-$222.06\\
		\hline
        $PC_1$ &    57.90 &    $-$1.12 &    1.88 &   $-$11.11\\
        $PC_2$ &    $-$5.88 &   $-$12.00 &    4.62 &    14.20\\
        $PE_1^u$ &   0.39 &    $-$9.47 &   $-$3.65 &     8.04\\
        $PB$ &         0.85 &    $-$1.06 &   $-$1.46 &    $-$0.53\\
        $PE_1^d+PE_2$ &   $-$0.55 &    $-$3.83 &     1.37 &    $-$0.31\\%\hline
        Penguin $\mathcal{P}$ &    52.71 &   $-$27.49 &     2.77 &    10.28\\
		\hline
		\hline
        \multicolumn{5}{c}{$\Lambda_b\to pK^-$}\\
        \hline
        $T$ &   853.60 &   $-$52.08 &  1190.21 &  $-$340.84\\
        $E_2$ & $-$66.28 &   $-$59.48 &   $-$50.31 &    79.56\\%\hline
        Tree $\mathcal{T}$ &   787.31 &  $-$111.55 &  1139.90 &  $-$261.28\\
		\hline
        $PC_1$ &    75.64 &    $-$0.82 &   $-$4.35 &   $-$13.81\\
        $PE_1^u$ &   0.10 &   $-$11.80 &   $-$4.76 &     9.93\\
        $PE_1^d$ &  $-$1.50 &    $-$7.38 &    1.66 &     2.09\\%\hline
        Penguin $\mathcal{P}$ &      74.23 &   $-$20.00 &   $-$7.45 &    $-$1.79\\
        \hline\hline
\end{tabular}
\end{table}

\begin{table*}[t]
    \footnotesize
	\centering
	\renewcommand{\arraystretch}{1.6}
	\begin{tabular*}{170mm}{c@{\extracolsep{\fill}}cccccc}
		\hline
        \hline
		 & $A_{CP}^{\rm dir}$ & $A_{CP}^{S\text{-wave}}(\kappa_S)$ & $A_{CP}^{P\text{-wave}}(\kappa_P)$ & $A_{CP}^\alpha$ & $A_{CP}^\beta$ & $A_{CP}^\gamma$\\%\text{-wave}
		\hline
		$\Lambda_b\to p\pi^-$ & $0.05^{+0.02}_{-0.03}$ & $0.17^{+0.05}_{-0.09}~(49\%)$ & $-0.06^{+0.04}_{-0.05}~(51\%)$ & $0.02^{+0.01}_{-0.02}$ & $0.22^{+0.08}_{-0.05}$ & $0.11^{+0.05}_{-0.06}$ \\
		$\Lambda_b \to  pK^-$ & $-0.06^{+0.03}_{-0.02}$ & $-0.05^{+0.05}_{-0.04}~(94\%)$ & $-0.21^{+0.39}_{-0.46}~(6\%)$ & $0.04^{+0.03}_{-0.04}$ & $-0.44^{+0.08}_{-0.04}$ & $0.02^{+0.06}_{-0.05}$ \\
		\hline\hline
		 & $A_{CP}^{\rm dir}$ & $A_{CP}^{S^T\text{-wave}}(\kappa_{S^T})$ & $A_{CP}^{(D+S^L)\text{-wave}}(\kappa_{D+S^L})$ & $A_{CP}^{P_1\text{-wave}}(\kappa_{P_1})$ & $A_{CP}^{P_2\text{-wave}}(\kappa_{P_2})$ & $A_{CP}^{\mathcal{J}}$\\
		\hline
		$\Lambda_b\to p \rho^-$ & $0.03^{+0.03}_{-0.05}$ & $0.01^{+0.01}_{-0.04}~(7\%)$ & $0.02^{+0.07}_{-0.03}~(44\%)$ & $0.03^{+0.04}_{-0.12}~(45\%)$ & $0.17^{+0.04}_{-0.06}~(4\%)$ & $-0.01^{+0.01}_{-0.01}$ \\
		$\Lambda_b\to pK^{\ast -}$ & $-0.05^{+0.10}_{-0.16}$ & $-0.15^{+0.12}_{-0.06}~(6\%)$ & $0.27^{+0.09}_{-0.27}~(33\%)$ & $-0.23^{+0.10}_{-0.18}~(55\%)$ & $-0.14^{+0.02}_{-0.10}~(6\%)$ & $0.02^{+0.04}_{-0.05}$ \\
		\hline\hline
		 & $A_{CP}^{\rm dir}$ & $A_{CP}^{S^T\text{-wave}}(\kappa_{S^T})$ & $A_{CP}^{(D+S^L)\text{-wave}}(\kappa_{D+S^L})$ & $A_{CP}^{P_1\text{-wave}}(\kappa_{P_1})$ & $A_{CP}^{P_2\text{-wave}}(\kappa_{P_2})$ & $A_{CP}^{UD}$\\
		\hline
		$\Lambda_b\to p a_1^-(1260)$ & $-0.01^{+0.04}_{-0.03}$ & $-0.22^{+0.10}_{-0.10}~(6\%)$ & $-0.11^{+0.03}_{-0.07}~(46\%)$ & $0.18^{+0.11}_{-0.06}~(40\%)$ & $-0.24^{+0.07}_{-0.13}~(8\%)$ & $-0.24^{+0.08}_{-0.13}$ \\
		$\Lambda_b\to p K_1^-(1270)$ & \multirow{2}*{$0.09^{+0.08}_{-0.05}$} & \multirow{2}*{$0.34^{+0.02}_{-0.06}~(8\%)$} & \multirow{2}*{$-0.11^{+0.12}_{-0.08}~(42\%)$} & \multirow{2}*{$0.19^{+0.17}_{-0.15}~(42\%)$} & \multirow{2}*{$0.33^{+0.04}_{-0.05}~(8\%)$} & \multirow{2}*{$0.26^{+0.04}_{-0.10}$} \\
        $(\theta_K=30^\circ)$ & ~ & ~ & ~ & ~ & ~ & ~ \\
		$\Lambda_b\to p K_1^-(1270)$ & \multirow{2}*{$0.07^{+0.05}_{-0.06}$} & \multirow{2}*{$0.46^{+0.02}_{-0.09}~(9\%)$} & \multirow{2}*{$0.06^{+0.11}_{-0.08}~(37\%)$} & \multirow{2}*{$-0.07^{+0.09}_{-0.10}~(45\%)$} & \multirow{2}*{$0.46^{+0.06}_{-0.07}~(9\%)$} & \multirow{2}*{$0.40^{+0.04}_{-0.09}$} \\
        $(\theta_K=60^\circ)$ & ~ & ~ & ~ & ~ & ~ & ~ \\
		\hline
		\hline
	\end{tabular*}
	\caption{CPV observables associated with (quasi-)two-body hadronic $\Lambda_b$ decays. The percentages in the parentheses indicate the proportions of partial-wave CPVs to the direct CPVs.}
	\label{tab:observables}
\end{table*}

The obtained direct CPVs and partial-wave CPVs of the $\Lambda_b\to p\pi^-$, $pK^-$ decays are presented in Table.~\ref{tab:observables}, with $\kappa_{S,P}$, shown as the percentages in parentheses, being the proportions of $S$- and $P$-wave contributions to the direct CPV.
Theoretical uncertainties arise mainly from the variation of the parameters involved in the DAs. 
It is noticed that the partial-wave CPVs can exceed $10\%$, i.e., $A_{CP}^{S\text{-wave}}(\Lambda_b\to p\pi)=0.17$ and $A_{CP}^{P\text{-wave}}(\Lambda_b\to pK)=-0.21$, similar to those in $B$ meson decays. 
However, the destruction between the $S$-wave CPV ($0.17$) and the $P$-wave CPV ($-0.06$) in $\Lambda_b\to p\pi^-$ reduces the magnitude of direct CPV to $0.05^{+0.02}_{-0.03}$, which agrees with the data in Eq.~(\ref{eq:AcpExp}). 
This demonstrates the previous discussion on the cancellation of partial-wave CPVs using topological diagrams. 
In the penguin-dominated decay of $\Lambda_b\to pK^-$, the $S$-wave CPV is small ($-$0.05) due to the small strong phase $\delta^{S\text{-wave}}=\delta_\mathcal{P}^S-\delta_\mathcal{T}^S=-7^\circ$, while the $P$-wave CPV is highly suppressed by the proportion $\kappa_P=6\%$ due to $|P|\ll |S|$. Hence, its direct CPV ($-0.06^{+0.03}_{-0.02}$) reveals no obvious deviation from zero either.

The large partial-wave CPVs can be probed by measuring the decay asymmetry parameters $\alpha$, $\beta$, and $\gamma$, which were first proposed by Lee and Yang \cite{Lee:1957qs}. These observables, related to the partial-wave amplitudes via $\alpha\propto \text{Re}(S^*P)$, $\beta\propto \text{Im}(S^*P)$ and $\gamma\propto |S|^2-|P|^2$,
are also predicted and listed in Table~\ref{tab:observables}.
It is found that the $\beta$-induced CPVs are significant but difficult to observe in current experiments; it is hard to determine experimentally the polarization of a final-state proton.

\begin{figure*}[tbhp]
	\centering
	\includegraphics[width=0.85\linewidth]{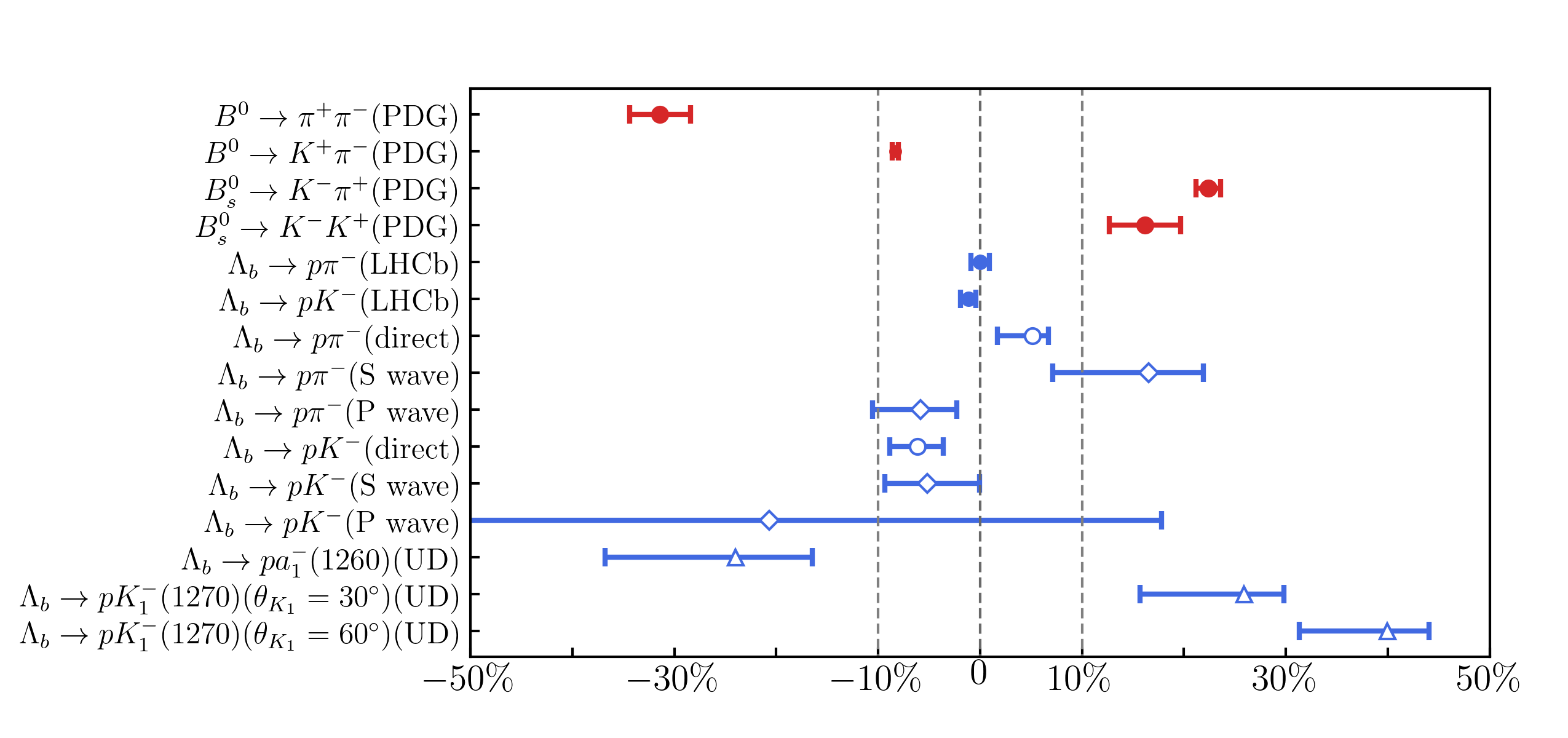}
	\caption{CPVs measured in $B$ meson and $\Lambda_b\to p\pi^-,pK^-$ decays, and our predictions.}
	\label{fig:EXPandPQCD}
\end{figure*}

%%%%%%%%%%%%%%%%%%%%%%%%%%%%%%%%%%
%The theoretical foundation of large partial-wave CPVs and their cancellations, which differentiate $b$-baryon decays from $B$-meson ones, is the main highlight of this work. 

\textit{$CP$ violations in $\Lambda_b\to pA$, $pV$}\textemdash{} 
The large partial-wave CPVs can certainly exist in other hadronic $\Lambda_b$ baryon decays. 
We further examine the modes $\Lambda_b\to pA$ with the axial-vector meson $A=a_1^-(1260)$ or $K_1^-(1270)$, and $\Lambda_b\to pV$ with the vector meson $V=\rho$ or $K^{*-}$. 
They share the same topological diagrams with $\Lambda_b\to p\pi^-,pK^-$, but with distinct meson DAs.

%\begin{equation}
%    \begin{split}
%        A_{CP}^{S^T} =& \frac{|S^T|^2-|\bar{S}^T|^2}{|S^T|^2+|\bar{S}^T|^2},\\
%        A_{CP}^{D+S^L} =& \frac{|D+S^L|^2-|\bar{D}+\bar{S}^L|^2}{|D+S^L|^2+|\bar{D}+\bar{S}^L|^2},\\
%        A_{CP}^{P_1} =& \frac{|P_1|^2-|\bar{P}_1|^2}{|P_1|^2+|\bar{P}_1|^2},\\
%        A_{CP}^{P_2} =& \frac{|P_2|^2-|\bar{P}_2|^2}{|P_2|^2+|\bar{P}_2|^2}.
%    \end{split}
%\end{equation}

%like Eq.~(\ref{eq:ACP=S+P})
%{\small
%\begin{equation}
%    \begin{split}
%        A_{CP}^{dir} =& \frac{2p_c(E_p+m_p)}{4\pi m_{\Lambda_b}(\Gamma + \bar{\Gamma})} \Big\{\frac{2|S^T|^2}{1+A_{CP}^{S^T}}A_{CP}^{S^T} + \frac{2|P_2|^2}{1+A_{CP}^{P_2}}A_{CP}^{P_2} \\
%        & + \frac{E_h^2}{m_h^2}\frac{|D+S^L|^2}{1+A_{CP}^{D+S^L}}A_{CP}^{D+S^L} + \frac{E_h^2}{m_h^2}\frac{|P_1|^2}{1+A_{CP}^{P_1}}A_{CP}^{P_1} \Big\},
%    \end{split}
%\end{equation}}
%or can be simplified into the following form according to the first order of approximation
%approximated like Eq.~\ref{eq:ACP=S+P},

The longitudinal and transverse polarized $\Lambda_b\to pA$ decay amplitudes are parametrized as 
\begin{equation}
\begin{split}
    \mathcal{A}^L(\Lambda_b\to pA)=&\bar{u}_p\epsilon_{L \mu}^{\ast} \Big(A_1^L\gamma^\mu\gamma_5+A_2^L\frac{p_p^\mu}{m_{\Lambda_b}}\gamma_5\\
    &+B_1^L\gamma^\mu+B_2^L\frac{p_p^\mu}{m_{\Lambda_b}} \Big) u_{\Lambda_b},\\
    \mathcal{A}^T (\Lambda_b\to pA)=&\bar{u}_p\epsilon_{T\mu}^{\ast }( A_1^T\gamma^\mu\gamma_5+B_1^T\gamma^\mu ) u_{\Lambda_b},
\end{split}\label{par}
\end{equation}
where $\epsilon_{L,T}$ are the longitudinal and transverse polarization vectors of the axial-vector meson, and the polarization amplitudes $A_1^{L,T}$, $A_2^L$, $B_1^{L,T}$, and $B_2^L$ form the partial-wave amplitudes $S^L=-A_1^L$, $S^T=-A_1^T$, $P_1 \approx -2B_1^L - B_2^L$, $P_2 \approx B_1^T$, and $D \approx -A_1^L + A_2^L$. 
The corresponding partial-wave CPVs are defined as in Eq.~(\ref{eq:partialCPV}), in terms of which the direct CPVs can be evaluated:
\begin{equation}
    A_{CP}^{dir}\approx  \kappa_{S^T}A_{CP}^{S^T}  + \kappa_{P_1}A_{CP}^{P_1}+ \kappa_{P_2}A_{CP}^{P_2} + \kappa_{D+S^L}A_{CP}^{D+S^L},
\end{equation}
with $\kappa_{S^T}=2|S^T|^2/\Pi$, $\kappa_{P_1}=(E_h^2|P_1|^2)/(m_h^2\Pi)$, $\kappa_{P_2}=2|P_2|^2/\Pi$, $\kappa_{D+S^L}=(E_h^2|D+S^L|^2)/(m_h^2\Pi)$ and  $\Pi \equiv 2|S^T|^2 + 2|P_2|^2 + E_h^2/m_h^2|D+S^L|^2 + E_h^2/m_h^2|P_1|^2$.
The parametrization in Eq.~(\ref{par}) and the above formulation also apply to the $\Lambda_b\to pV$ modes.

%Notice the additional $D$-wave amplitude.

The predictions for the CPVs in the  $\Lambda_b\to pV$, $pA$ decays are collected in Table.~\ref{tab:observables}. 
As the $K_1(1270)$-$K_1(1400)$ mixing angle $\theta_K$ is not yet well determined \cite{Shi:2023kiy}, we take the typical values $\theta_K=30^\circ$ and $60^\circ$ for illustration. Most of the results are not very sensitive to $\theta_K$. 
It is found that the $D+S^L$ and $P_1$ components dominate the direct CPVs because of the relativistic enhancement factor $E_h/m_h$.
The partial-wave CPVs of the $\Lambda_b\to pV$, $pA$ decays also exceed $10\%$;
the $P_2\text{-wave}$ CPVs of $\Lambda_b\to p\rho^-$ and $pa_1^-(1260)$ are $17\%$ and $-24\%$, respectively;
the $(D+S^L)\text{-wave}$ CPV of $\Lambda_b\to pK^{*-}$ is $27\%$;
the partial-wave CPVs of $\Lambda_b \to pK^-_1(1270)$ even reaches the order of $30\%$.
However, the direct CPVs of these modes are all small.
The relative sign of the partial-wave amplitudes involved in the $\Lambda_b\to pK^{*-}$, $pa_1^-(1260)$ and $pK_1(1270)$ decays can be argued in the same manner. Their tiny direct CPVs  trace back to the strong cancellation of the major $(D+S^L)$- and $P_1\text{-wave}$ CPVs, similar to the  $\Lambda_b\to p\pi^-$ case.

Motivated by the sizable partial-wave CPVs, we construct promising observables associated with angular distributions of final states in multibody $\Lambda_b$ decays, which can be measured in future experiments.
With $\theta$ as the angle between the normal of the $A \to h_1h_2h_3$ decay plane and the momentum of $A$ in the $\Lambda_b$ rest frame, the angular distribution for the $\Lambda_b\to pA\to ph_1h_2h_3$ decay reads
\begin{equation}\label{eq:angular-pA}
\begin{split}
    \frac{d\Gamma}{d\cos\theta}\supset R~
    %\left\{ m_+m_-\frac{E_p+m_p}{p_c} \right\}
    \mathcal{R}e(S^TP_2^\ast)~\cos\theta.
    % (\Lambda_b\to pA\to ph_1h_2h_3) \propto  \\
    % & 4\left\{\Big|m_+S^T\Big|^2+\Big|m_-\frac{E_p+m_p}{p_c}P_2\Big|^2\right\}\frac{1+cos^2\theta}{2}\\
    % & +2\Big\{ \Big|\frac{m_+(m_{\Lambda_b}-m_p)+m_-p_c}{m_h}S^L + \frac{m_-E_h(E_p+m_p)}{m_hp_c}D\Big|^2\\
    % & + \Big|\frac{m_+E_h}{m_h}P_1\Big|^2 \Big\}sin^2\theta\\
    % & -8\frac{R^-}{R^+}\left\{ m_+m_-\frac{E_p+m_p}{p_c}Re(S^TP_2^\ast) \right\}cos\theta,
\end{split}
\end{equation}
The nonperturbative factor $R$ parametrizes the effect and kinematics of the strong decay $A\to h_1h_2h_3$~\cite{Berman:1965gi}, which will be canceled in the proposed CPV observable below.
%$R$ is nonperturbative and difficult to be calculated, but can be cancelled in the definition of CP violation in the following. 
%$m_\pm=\sqrt{(m_{\Lambda_b}\pm m_p)^2-m_A^2}$ with $m_A$ being mass of the axial vector $A$.
The full expression of the angular distribution is referred to \cite{Wang:2024qff}, which contains additional pieces proportional to $1$ and $\cos^{2}\theta$.
We consider the up-down asymmetry for four-body decays:
{\small
\begin{equation}
\begin{split}
&A_{UD}\equiv\frac{\Gamma(\cos\theta>0) - \Gamma(\cos\theta<0)}{\Gamma(\cos\theta>0) + \Gamma(\cos\theta<0)} =   R ~ \mathcal{R}e(S^TP_2^\ast),
\end{split}
\end{equation}} 
from which the $R$-independent CPV is defined:
\begin{equation}
\begin{split}
A_{CP}^{UD}&=\frac{A_{UD}+\bar{A}_{UD}}{A_{UD}-\bar{A}_{UD}},\\
% & \propto \frac{Re(S^TP_2^\ast+\bar{S}^T\bar{P}_2^\ast)}{Re(S^TP_2^\ast-\bar{S}^T\bar{P}_2^\ast)},
\end{split}
\end{equation}
with $\bar{A}_{UD}$ denoting the charge conjugate of $A_{UD}$.
Note that this CPV is induced by the partial-wave amplitudes $S^T$ and $P_2$. 
The angular distribution in the $\Lambda_b\to pV\to ph_1h_2$ decay is described by $1+\mathcal{J}(3\cos^2\theta -1)/2$ \cite{Wang:2024qff}, where $\theta$ is the angle between $h_1$ in the $V$ rest frame and the momentum of $V$ in the $\Lambda_b$ rest frame.
Likewise, the asymmetry $A_{CP}^{\mathcal{J}} = (\mathcal{J}-\bar{\mathcal{J}})/(\mathcal{J}+\bar{\mathcal{J}})$ for $\Lambda_b \to pV$ can be defined.

%up-down asymmetry is more convenient for the following form in which the unknown $R$ is offsetted,

%\begin{figure}[tbhp]
%	\centering
%	\subfigure{
%		\begin{minipage}[]{0.46\linewidth}
%			\includegraphics[width=0.95\linewidth]{figures/angle-distri-pV.png}
%		\end{minipage}
%	}
%	\subfigure{
%		\begin{minipage}[]{0.46\linewidth}
%			\includegraphics[width=0.95\linewidth]{figures/angle-distri-pA.png}
%		\end{minipage}
%	}
%   \caption{Kinematics for the $\Lambda_b\to pV\to ph_1h_2$(left panel) and $\Lambda_b\to pA\to ph_1h_2h_3$(right panel) decays.}
%	\label{fig:angle-distri}
%\end{figure}

The predictions for $A_{CP}^{UD}$ are given in Table.~\ref{tab:observables},  
which exceed $20\%$ in the $\Lambda_b \to pA$ decays owing to large strong phase difference between the $S$- and $P$-wave amplitudes.
Hence, $A_{CP}^{UD}$ with controllable uncertainties serves as an ideal observable for establishing CPVs in  $\Lambda_b$ decays experimentally.
%Those for $A_{CP}^{\mathcal{J}}$ diminish as indicated in Table.~\ref{tab:observables}.
The measurements of $A_{CP}^{UD}$ are based on the four-body decays $\Lambda_b\to  p \pi^+ \pi^- \pi^-$ and $p K^- \pi^+ \pi^-$ for $pa_1(1260)$ and $pK_1(1270)$, respectively, both of which have large data samples at LHCb.
The LHCb has reported the direct CPVs of the above modes from the Run 1 data corresponding to the integrated luminosity 3 fb$^{-1}$ at the center-of-mass energies $\sqrt{s}=$ 7 and 8 TeV \cite{LHCb:2019jyj}. The sum of particle and antiparticle channels gives the signal yields of $\Lambda_b\to pa_1$ and $pK_1$ around 800 and 1000, respectively. 
The yields would be enhanced by 4 times with the Run 2 data from twice the integrated luminosity of 6 fb$^{-1}$ and twice the cross section at a higher collision energy of $\sqrt{s}=13$ TeV.
According to the plan with 50 fb$^{-1}$ at the end of Run 4 \cite{Chen:2021ftn}, the signal yields of Run 3 + 4 would be greater than those of Run 2 by a factor of $(50-9)$ fb$^{-1}/6$ fb$^{-1}\times 2\sim14$, where the factor of 2 comes from the enhancement of trigger efficiencies starting from Run 3 \cite{TriggerEfficiency}.
The total events with the full Run 1+2+3+4 data are expected to be a factor of 60 compared to the Run 1 data, as large as $48~000$ and $60~000$ for $\Lambda_b\to pa_1$ and $pK_1$, respectively.
With $A_{UD}[\Lambda_b\to p a_1]=-0.09\pm0.02$, $A_{UD}[\Lambda_b\to p K_1(\theta_K=30^\circ)]=-0.19\pm0.04$, and $A_{UD}[\Lambda_b\to p K_1(\theta_K=60^\circ)]=-0.24^{+0.06}_{-0.05}$, the statistical uncertainties of $A_{CP}^{UD}$ are then roughly $5\%$ for $\Lambda_b\to p a_1$ and $4\%$ for $\Lambda_b\to p K_1$ at the end of LHCb Run 4. 
The systematic uncertainties of $A_{CP}^{UD}$ are smaller based on the control sample of $\Lambda_b^0\to \Lambda_c^+(\to pK^-\pi^+)\pi^-$ as reported in \cite{LHCb:2019oke}. 
Therefore, the up-down asymmetric CPVs in $\Lambda_b\to p a_1$ and $p K_1$ of the order of $20\%- 40\%$ have a large possibility of being identified in the near future. Such measurements could establish $CP$ violations in baryon decays.

%%%%%%%%%%%%%%%%%%%%%%%%%%%%%%%%%%
\textit{Conclusion}\textemdash{} 
The rich data samples and complicated dynamics in multibody bottom baryon decays offer high prospects for exploring baryon CPVs. This Letter presented the first full QCD analysis on two-body hadronic $\Lambda_b$ decays in the PQCD approach. Our predictions for the total and partial-wave CPVs of the considered modes are summarized and compared with the available data for bottom hadron decays in Fig.~\ref{fig:EXPandPQCD}. The study elucidates the mechanism responsible for the measured small CPVs in $\Lambda_b\to p\pi^-,pK^-$, in contrast to the sizable CPVs in the similar $B$ meson decays. The partial-wave CPVs in $\Lambda_b\to p\pi^-$ reach $10\%$ potentially, but the destruction
between them leads to the tiny CPV. The direct CPV of $\Lambda_b\to pK^-$ is primarily attributed to the modest $S$-wave CPV.
We have also extended our investigation to the CPVs in the modes with vector and axial-vector final states. 
Our work suggests that certain partial-wave CPVs in bottom baryon decays, especially those related to angular distributions, can be sufficiently significant, and probed experimentally to establish baryon CPVs. It opens up avenues for deeply understanding the dynamics in heavy baryon decays and their CPVs.

\textit{Note added}\textemdash
The LHCb Collaboration has reported the first observation of the direct CPV in $\Lambda_b$ baryon decays \cite{LHCb:2025ray} recently. This discovery undoubtedly marks a significant milestone in the relevant fields.

%%%%%%%%%%%%%%%%%%%%%%%%%%%%%%%%%%
\textit{Acknowledgement}\textemdash  
We acknowledge Jun Hua, Yan-Qing Ma, Ding-Yu Shao, and Jian Wang, for their helpful suggestions on the numerical calculations, and Pei-Rong Li and Yan-Xi Zhang for their valuable comments on the experimental measurements. This work was supported in part by Natural Science Foundation of China under
Grant No. 12335003, the Fundamental Research Funds for the Central Universities under No. lzujbky-2023-stlt01 and No. lzujbky-2024-oy02, and the Super Computing Center at Lanzhou University.

%\begin{equation}
%\begin{split}
%    H_{\frac{1}{2},1} =& \sqrt{2}M_+ S^T -\sqrt{2}M_- \frac{E_p+m_p}{p_c} P_2,\\
%    H_{-\frac{1}{2},-1} =& -\sqrt{2}M_+ S^T -\sqrt{2}M_- \frac{E_p+m_p}{p_c} P_2,\\
%    H_{\frac{1}{2},0} =& -\frac{M_+(m_{\Lambda_b}-m_p)+M_-p_c}{m_h}S^L\\
%    &- \frac{M_-E_h(E_p+m_p)}{m_hp_c}D - \frac{M_+E_h}{m_h}P_1,\\
%    H_{-\frac{1}{2},0} =& \frac{M_+(m_{\Lambda_b}-m_p)+M_-p_c}{m_h}S^L\\
%    &+ \frac{M_-E_h(E_p+m_p)}{m_hp_c}D - \frac{M_+E_h}{m_h}P_1,
%\end{split}
%\end{equation}

%%%%%%%%%%%%%%%%%%%%%%%%%%%%%%%%%%
%\input{appendix}

%%%%%%%%%%%%%%%%%%%%%%%%%%%%%%%%%%

\end{document}